\documentclass[a4paper]{article}

\usepackage[english]{babel}
\usepackage[utf8x]{inputenc}  
\usepackage[T1]{fontenc}
\usepackage{stmaryrd} 

\usepackage{amsmath}
\usepackage{amsfonts}
\usepackage{amssymb}
\usepackage{graphicx}
\usepackage{url}

\usepackage[colorinlistoftodos]{todonotes}
\usepackage[allcolors=blue]{hyperref}

\begin{document}
\pagenumbering{gobble}

\Large
 \begin{center}
Intelligent Drone Swarm for Search and Rescue Operations at Sea\\ 

\hspace{10pt}

\large
Vincenzo Lomonaco$^1$, Angelo Trotta$^1$, Marta Ziosi$^2$, Juan de Dios Yáñez Ávila$^3$, Natalia Díaz-Rodríguez$^4$ \\

\hspace{10pt}

\small     
$^1$University of Bologna, Italy. $^2$The London School of Economics and Political Science, UK. 
$^3$Waredrone \& Higher Technical School of Engineering, University of Seville, Spain. 
$^4$ENSTA ParisTech, France.\\
vincenzo.lomonaco@unibo.it, angelo.trotta5@unibo.it, m.ziosi@lse.ac.uk, juandediosyanez@waredrone.es, natalia.diaz@ensta-paristech.fr

\end{center}

\hspace{10pt}

\normalsize





\begin{figure}
\centering
  \includegraphics[width=0.67\linewidth]{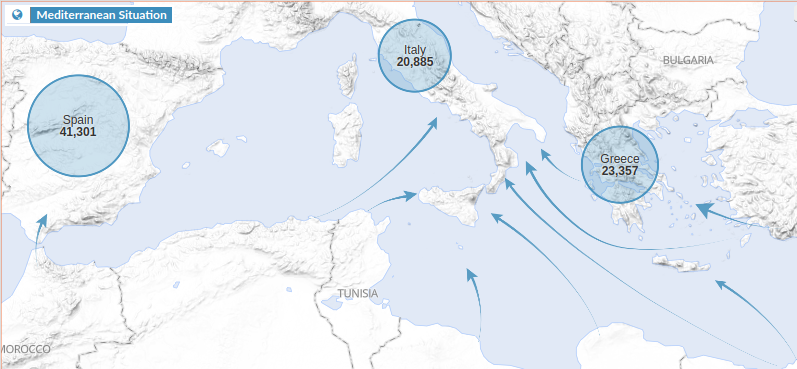}
  \caption{\footnotesize Migration flows in the Mediterranean area as documented by the UNHCR.
  Numbers for each nation indicate the total number of arrivals in 2018 (last update 2018-10-01).}
  \label{fig:a}
\end{figure}

The United Nations Secretary General Strategy on New Technologies reads \emph{``The United Nations system will support the use of new technologies like artificial intelligence} [...] \emph{and robotics to accelerate the achievement of the 2030 Sustainable Development Agenda''} \cite{UN}.  The idea of this project is to use Artificial Intelligence (AI) technology (specifically, an intelligent drone swarm) for search and rescue (SAR) operations at sea. The relevance of the drone swarm to the call for the use of technology is clear. However, how are these SAR operations relevant to the Sustainable Development Goals' (SDGs) agenda? In this project, we refer to SAR operations specifically in the context of the migrant crisis in the Mediterranean sea. In this context, SAR operations are relevant to the SDGs on three different levels. Firstly, Paragraph 23 of the Agenda states: \emph{``Those whose needs are reflected in the Agenda include all children, youth,} [...] \emph{refugees and internally displaced persons and migrants''} \cite{2030}. 
Secondly, among the goals we find: no poverty, good health and well-being, decent work and economic growth, reduced inequalities, peace, justice and strong institutions \cite{sust}. 
The completion of these goals is unattainable without an improvement in the migrants situation. 
Thirdly, among the 169 SDGs' specific targets, we find \emph{``facilitate orderly, safe, regular and responsible migration''} 
\cite{sust}.
The theoretical considerations above and the practical urgency of the migrant crisis establish the relevance of the topic of our proposal for SDGs.
In fact, in the last 5 years, more than 16 thousands people have lost their lives in the Mediterranean \cite{unhcr1} (see Fig. \ref{fig:a}). 

However, is the 
use of drones
apt for the problem at hand? 
Their use
in migrants SAR operations at sea has been occasionally documented. NGOs such as the Migrant Offshore Aid Station 
(MOAS) 
and Sea-Watch have recently partnered with tech start-ups to test the use of drones to rescue migrants \cite{ziagemma, seawatch}. On one hand, these cases point at single NGOs, rather than to a more wide-spread adoption. On the other, many technical, financial and legal problems have remained unresolved and have impeded a successful implementation. 


Monitoring a large area of sea for search and rescue operations is a difficult task \cite{Wong17, breivik13}. Current solutions involve the use of ships and highly expensive aerial vehicles piloted by human experts who are also in charge of swiftly detecting emergency situations. More recently, research in this context focused on the use of unmanned vehicles for reducing operational costs and providing more efficient support for SAR operations.
These vehicles can operate at different levels: underwater \cite{Venkatesan16, Matos16}, on the surface \cite{Mendonca16, Matos13} and above it \cite{yeong15, Topputo15}.
Focusing on the unmanned aerial vehicles (UAVs), even though they are generally faster and more efficient in covering large areas, 
they lack in flight autonomy and communication capacity due to the limited Line of Sight (LoS) in the ground-to-air communication link.
For example, a common scenario for SAR operations would be to monitor a large area of $30\times30\ km^2$ at a distance of $20 km$ from the base (ship or land station). This kind of operations are not easily accomplished with a single UAV. In fact, with the expected battery lifetime (of about $2h$) and with an average speed of $50 mph$ of an off-the-shelf fixed-wing UAV, covering an area of more than $10\times10\  km^2$ is not feasible nowadays. Moreover, the wireless communication link between the UAV and the base station, which is used for video streaming and for the actual detection of endangered people, is not strong enough to support the 
bandwidth requirement
after a few kilometers.

We propose an \emph{Intelligent Drone Swarm} for tackling the aforementioned problems. The central idea is to use an autonomous fleet of intelligent UAVs to accomplish the task. 
The self-organizing UAVs network would enable the coverage of a larger area in less time, reducing the impact of the limited battery capacity. 
Recent works on UAVs networks shown their ability of coverage and connectivity improvements in emergency scenarios \cite{trotta15}.
Indeed, the UAVs swarm is capable of generating a multi-hop communication network that will guarantee a higher bandwidth due to aerial-to-aerial communication links provided by longer LoS connections.
Artificial Intelligence tools, in this case, would greatly improve not only the autonomy and coordination of the UAVs network, but also the communication efficiency.
On one hand, autonomy and coordination are improved due to the autonomous adaptation of mobility actions and communication parameters to non-stationary environmental conditions. On the other, communication efficiency is ensured with the usage of smart detection algorithms locally running on the UAV, which limit the amount of information sent to the base station. Recent advances in AI techniques, such as specialized AI chips \cite{Gokhale17} and quantization/pruning techniques \cite{han15}, may also help reducing the energy and computational resources needed for detection.


Even though the costs will be high upfront, the scalability and sustainability of this system will ensure a proportional reduction of any additional long term expense. Nevertheless, many challenges will arise during the deployment of this kind of networks in real-world scenarios: the hostility of the sea environment with strong winds and the unavailability of safe landing spots may constitute an hard impediment for current state-of-the-art UAVs. However, we believe that a research effort in this direction is paramount for search and rescue operations at sea with the ultimate goal of saving human lives.

\newpage

\bibliography{library}{}
\bibliographystyle{unsrt}


\end{document}